\documentstyle[11pt]{article}

\begin{document}

\centerline{\bf\large Contributions from Goldstone-boson-exchange
to baryon spectra in the MIT Bag Model} \vspace{1cm}

\centerline{Da-Heng He$^1$, Yi-Bing Ding$^2$, Xue-Qian Li$^1$ and
Peng-Nian Shen$^3$}

\vspace{0.3cm}

1. Department of Physics, Nankai University, Tianjin, 300071,
China.

2. Institute of Physics, Graduate School of Chinese Academy of
Sciences, Beijing, 100049, China.

3. Institute of High Energy Physics, P.O.Box 918-4, Beijing,
100049, China.

\vspace{1cm}

\begin{center}
\begin{minipage}{13cm}
\noindent Abstract:

We discuss contributions of chiral bosons to baryon spectra in the
MIT bag model. It is believed that within hadrons, chiral bosons
are degrees of freedom which are independent of gluons to provide
strong interactions between quarks. In the original MIT bag model,
only interaction mediated by gluon exchanges was considered, by
contrast, in this work we take into account the interaction
mediated by the exchanges of chiral bosons $\sigma$ and
$\pi^{(\pm,0)}$. Then following the standard approach, we minimize
the effective hamiltonian which includes both the contributions
from gluon and chiral-boson exchanges with respect to the bag
radius to obtain the effective radius. By re-fitting the spectra
of baryons, we find that the contributions from the boson-exchange
may be 40\% of that from gluon-exchanges and meanwhile the bag
constant $B$, the zero-point energy $z_0$ almost do not change. It
indicates that in the original version of the MIT bag model, the
intermediate-distance interaction due to the chiral-boson
exchanges is attributed into the effective coupling $\alpha_c$
which stood for the short-distance interaction caused by the gluon
exchanges and the long-distance effects reflected by $B$ and $z_0$
are not influenced.

\end{minipage}
\end{center}

\vspace{2cm} \baselineskip 22pt

\noindent {I. Introduction}\\

It is generally believed that QCD  is the successful theory for
strong interaction and nowadays, nobody ever doubts its validity.
Due to the asymptotic freedom, at higher energy processes, all the
physical quantities, such as cross sections, can be calculated
perturbatively and the results are very accurate. However, when we
deal with the hadron physics, the typical energy scale is
$\Lambda_{QCD}\sim 200$ MeV, at this region, the non-perturbative
QCD effects dominate and any perturbative QCD calculations become
questionable. So far, there are no reliable ways to properly
handle the non-perturbative QCD based on any underlying
principles.

To evaluate the hadron spectra and their hyperfine structure etc.
one needs to invoke some concrete models which may implement the
non-perturbative behaviors of QCD into the models and concerned
parameters. The traditional methods include the potential model,
MIT bag model and many others. In all the models, the
short-distance interaction between quarks is induced by exchanging
hard gluons and the leading order is the one-gluon exchange. But
the ways to describe the long-distance effects of QCD are
different for different models. For example, in the potential
model, a confinement term is phenomenologically introduced and the
concerned parameters must be obtained by fitting data. There are
several commonly adopted forms for the confinement term, and the
most common one is the linear potential which seems to be
consistent with the lattice results.

For the MIT bag model, a rigid bag-boundary which prevents outward
flux of quarks, replaces the linear potential to provide the
confinement. Inside the bag, quarks, at the zeroth order
approximation, are free of interactions, and obey the Dirac
equation for free fermion with a non-trivial boundary condition,
i.e. the outward flux is zero at the bag boundary. Then at the
next-to-leading order, one needs to take interactions among quarks
into account. As DeGrand et al.\cite{Jaffe} suggested, to this
approximation, the one-gluon exchange is responsible for the
interaction which can be expressed as couplings of the magnetic
dipole moments of quarks, obviously, it is equivalent to the
description where the interaction energy is achieved in terms of
the one-gluon-exchange mechanism according to the quantum field
theory. We have applied the method to evaluate the spectra of
baryons which contain two heavy quarks (b and/or c)\cite{He}.

When evaluate the energy caused by the effective interaction
between quarks, one needs to sandwich hamiltonian induced by the
one-gluon-exchange between the zeroth order wavefunctions of
quarks and calculate the expectation values. The expectation
values are the interaction energy between quarks. So far, the
whole procedure is perturbative, but later on, one needs to obtain
a new bag radius by minimizing the total energy which includes
both the zeroth order and newly derived next-to-leading order
contributions, with respect to the bag radius. Therefore the bag
model treatment is not totally perturbative.

Moreover, many research works indicate that in hadrons, not only
gluons and quarks, but also the chiral bosons, such as $\sigma$,
$\pi$ and even kaons, can be independent degrees of freedom
\cite{Win,Geor1,Geor2,yanmulin}. Namely, exchange of chiral bosons
is not included in exchange of hard multi-gluons. The gluons are
the QCD gauge bosons and possess color charges, so that they
interact among themselves. At lower energies, they cannot
propagate far, therefore can only be responsible for
short-distance interaction. If one demands that the one-gluon
exchange determines only the short-distance interaction, a
long-distance interaction which cannot be derived in the framework
of perturbation, is responsible for the confinement. Is the
picture too simplified? In other words, it should be asked if the
intermediate-distance interaction needs to be independently
evaluated, i.e. separated from both short- and long-distance
interactions as a distinct one.

Some authors suggest that in hadrons the asymptotic
freedom\cite{tho} completely applies and the  exchange of hard
gluons does not contribute to the spectra at all. Instead, only
the intermediate-distance and long-distance interactions
contribute. For example in the QMC Model\cite{Gui,Fle,Sai,Blu},
only chiral bosons ($\sigma,\pi,K$) and light vector bosons
($\rho,\omega,\phi$) are considered. It is interesting to notice
that the propagator of a chiral boson ${e^{-mr}\over r}$ seems to
be more suppressed at larger distance than the propagator of gluon
${1\over r}$. But it is in the perturbative sense. As a gauge
boson of Yang-Mills gauge field\cite{Yang}, gluons interact among
themselves and cannot propagate far, by contrary, the chiral boson
is color-singlet, so that does not suffer from this constraint.
The interaction induced by exchange of chiral bosons can be
considered as the intermediate-distance interaction.

It is natural to ask if one can omit any of the three kinds of
interactions which come from different aspects of QCD.
Phenomenologically, the question is if we can attribute any of the
interactions into the others by adjusting the concerned
parameters. By the literature, it is definitely plausible for
estimation of hadron spectra, but then the physics picture is not
complete and maybe, some hyperfine properties of hadrons would be
smeared away. Thus, we may wish to re-study the physics picture by
including all the three interactions and see if a complete physics
picture can be built up. Based on the commonly accepted
principles, we study the contributions from the short-distance
effects which are induced by the hard-gluon exchanges,
intermediate-distance contributions which are caused by the
exchanges of chiral bosons and the long-distance effects in a
unique framework. Namely, we investigate their respective
contributions in the MIT bag model where the long-distance effects
are provided by the bag-boundary and perhaps, also the zero-point
energy.

Our strategy in this work is that the effective hamiltonian which
accounts for contributions from both the hard-gluon exchange and
chiral-boson-exchange, is sandwiched between the zeroth order
wavefunctions of quarks to obtain a total energy, while the
long-distance effects are reflected in the bag-boundary condition
of the Dirac equation. The effective vertices between quark and
chiral bosons are described by the linear $\sigma$ model where the
$\sigma$ boson remains as an independent particle. After obtaining
the total energy, we minimize it with respect to the bag radius,
and the minimum is supposed to correspond to the hadron spectra.
Indeed, there is a zero-point energy which should be included and
determined by fitting data. It is natural to suppose that it is a
universal for all the baryons and can be fixed by experimental
data. The same problem exists in the potential model in fact.

More concretely, based on the principles and rules of quantum
field theory, we formulate the effective hamiltonian and evaluate
the contribution from the intermediate-distance interaction to the
total energy, which is caused by exchanging chiral bosons.

We also briefly discuss possible contributions of three-body
interactions. Namely it seems that the three constituent quarks
may interact via a three-gluon vertex, but a symmetry analysis
\cite{wang} indicates  that the net contribution is null due to
the color-singlet requirement for hadrons. Then we consider the
three-body intermediate-distance interaction via a
$\sigma-\pi-\pi$ coupling, since the corresponding structure is
very complicated, it is difficult to reach a complete solution.
Instead, we are going to estimate the order of magnitude of such
contribution. Only considering a simplified breathing mode which
is believed to be the leading mode, we qualitatively and
half-quantitatively evaluate its contribution and find that it is
much smaller than the two-body interaction and can be negligible
for practical computations.

This work is organized as follows. After this long introduction we
present the formulation for the interaction between quarks which
are caused by exchange of chiral mesons based on the principles
and Feynman rules of quantum field theory. In Sec.III, we present
the numerical results where the concerned parameters and inputs
are listed out explicitly. Then in Sec.IV, we discuss the
three-body interactions, derive the formulation of the effective
hamiltonian for the  breathing mode which is a simple and rough
approximation, we give our numerical estimates. The last section
is devoted to our conclusion and discussions.\\

\noindent{II. Formulation}

The short-distance interactions caused by gluon exchanges have
already been well formulated in the original works about the MIT
model\cite{Jaffe}. The color magnetic interaction energy is
written as
\begin{eqnarray}
\Delta
E_{m}=8\alpha_{c}\lambda\sum_{i>j}\frac{\mu(m_i,R)\mu(m_j,R)}{R^3}I(m_iR,m_jR)
(\overrightarrow{\sigma_i}\cdot\overrightarrow{\sigma_j}),
\end{eqnarray}
and the expressions of $\mu(m,R)$ and $I(m_iR,m_jR)$ are given in
\cite{Jaffe}. As DeGrand et al. proved, the color-electric
interaction energy is quite small and  we can simply ignore it.

The one-gluon-exchange results in a short-distance interaction and
$\Delta E_m$ is included in the total energy. As argued in
\cite{Win,Geor1,Geor2,yanmulin}, the chiral boson exchange could
be independent of the one-gluon-exchange and corresponds to the
intermediate-diatance interaction. Now let us turn to the
interacting energy which is caused by chiral boson-exchange.

The formulation to be used is directly derived from the principle
of quantum field theory\cite{He,Lif}:
\begin{eqnarray}
E_{int}=\int
\overline{\psi}'_1\Gamma_1\psi_1D_{prop}\overline{\psi}'_2\Gamma_2\psi_2d^3xd^3y,
\end{eqnarray}
where $\psi_1$, $\psi_2$ are the zeroth order bag wave functions
of two interacting quarks, $D_{prop}$ is the gauge boson
propagator in coordinate space and $\Gamma$ is the coupling
vertex.

First of all, for quark-meson coupling, the chiral lagrangian in
SU(2) is\cite{Mae}:
\begin{equation}
{\cal
 L}_{NP}=i\overline{\psi}(\gamma^\mu\partial_\mu-m)\psi+g\overline{\psi}{\mbox{\boldmath
$\Sigma$}}\psi,
\end{equation}
where
\begin{equation}
{\mbox{\boldmath $\Sigma$}}=\sigma+i\gamma_5{\mbox{\boldmath
$\tau$}}\cdot{\mbox{\boldmath $\pi$}},
\end{equation}
and $g$ is the effective coupling constant between scalar or
pseudoscalar meson and quarks in linear $\sigma$ model. To achieve
this effective coupling constant $g$, one may use the data of
$p-p$ scattering and the quark-hadron relation. However, in our
case, this method seems not work. Because for evaluating the
spectra, the quarks are confined and the effective coupling should
be close to the the value employed in the potential model which is
gained by fitting the baryon spectra. In fact, here we just keep
$g$ as a free parameter and determine it by fitting data and then
compare its value with that in the potential model and see if it
makes sense.

This SU(2) linear $\sigma$ model can be extended into SU(3) space
and applied to the calculations in the MIT bag model.

In SU(3) space, ${\mbox{\boldmath $\Sigma$}}$ is generalized as
\cite{Geo,Sch}:
\begin{eqnarray}
{\mbox{\boldmath $\Sigma$}}=\sigma+i\gamma_5T^a\phi^a_p,
\end{eqnarray}
and
\begin{eqnarray}
T^a\phi^a_p=\sqrt{2}\left (
\begin{array}{ccc}
\frac{\pi^0}{\sqrt{2}}+\frac{\eta}{\sqrt{6}} & \pi^+ & K^+\\
\pi^- & -\frac{\pi^0}{\sqrt{2}}+\frac{\eta}{\sqrt{6}} & K^0 \\
K^- & \overline{K^0} & -\sqrt{\frac{2}{3}}\eta
\end{array} \right ).
\end{eqnarray}
The propagator of chiral boson in the instantaneous approximation
i.e $q_0=0$ is:
\begin{eqnarray}
D_{prop}({\bf q})=\frac{1}{{\bf q^2}+m^2}.
\end{eqnarray}
By a Fourier-transformation, we can write it in the coordinate
space as:
\begin{eqnarray}
D_{prop}(r)&=&\int e^{i{\bf q}\cdot {\bf r}}\frac{1}{{\bf q^2}+m^2}\frac{d^3{\bf q}}{(2\pi)^3}\nonumber \\
&=& {}\frac{e^{-mr}}{r}.
\end{eqnarray}
where $r$ is the relative distance between two interacting quarks.
With these theoretical preparations, we will derive the
formulation of the interacting energy due to one-meson-exchange:
\begin{eqnarray}
E_{q\sigma q}&=&\int \overline{q}_1(x_1)gq_1(x_1)\frac{e^{-M_\sigma r}}{4\pi r}\overline{q}_2(x_2)gq_2(x_2)d^3x_1d^3x_2\nonumber \\
&=& {}\frac{g^2}{4\pi}N^2\int [j^2_0(\frac{\chi
x_1}{R})-j^2_1(\frac{\chi x_1}{R})]\frac{e^{-M_\sigma
r}}{r}[j^2_0(\frac{\chi x_2}{R})-j^2_1(\frac{\chi
x_2}{R})]d^3x_1d^3x_2,
\end{eqnarray}
and
\begin{eqnarray}
E_{qpq}&=&\sum_p\int \overline{q}_1(x_1)ig\gamma^5q_1(x_1)\frac{e^{-M_pr}}{4\pi r}\overline{q}_2(x_2)ig\gamma^5q_2(x_2)d^3x_1d^3x_2\nonumber \\
&=& {}\sum_p\frac{g^2}{4\pi}4N^2\int j_0(\frac{\chi
x_1}{R})j_1(\frac{\chi x_1}{R})({\mbox{\boldmath
$\sigma_1$}}\cdot{\bf x}_1)\frac{e^{-M_p r}}{r}\nonumber \\
&& j_0(\frac{\chi x_2}{R})j_1(\frac{\chi x_2}{R})({\mbox{\boldmath
$\sigma_2$}}\cdot{\bf x}_2)d^3x_1d^3x_2
\end{eqnarray}
where $E_{q\sigma q}$ is the interacting energy due to $\sigma$
meson exchange and $E_{qpq}$ is a sum of the contributions of
various pseudoscalar mesons. $q_1(x_1)$ and $q_2(x_2)$ are wave
functions of quarks; $N$ is the normalization constant;
$M_{\sigma}$, $M_p$ are the masses of the corresponding scalar and
pseudoscalar mesons.

In the calculation, one finds that the integration (13) (14) would
turn infinite when
\begin{eqnarray}
r=|{\bf x}_1-{\bf x}_2|\rightarrow 0.
\end{eqnarray}
No doubt, such infinity is only formal, if one can deal with a
whole computation properly, this divergency should disappear. To
avoid this problem, we introduce a new coordinate system:
\begin{eqnarray}
&&{\bf x}_1=\bf X,\nonumber \\
&&{\bf x}_1-{\bf x}_2=\bf Y.
\end{eqnarray}
Replacing coordinate ${\bf x}_1$, ${\bf x}_2$ by $\bf X$, $\bf Y$,
we have
\begin{eqnarray}
&E_{q\sigma q}&=\frac{g^2}{4\pi}N^2\int [j^2_0(\frac{\chi x_1}{R})-j^2_1(\frac{\chi x_1}{R})]d^3x_1\times\nonumber \\
&& {}\int\frac{e^{-M_\sigma |{\bf x}_1-{\bf x}_2|}}{|{\bf x}_1-{\bf x}_2|}[j^2_0(\frac{\chi x_2}{R})-j^2_1(\frac{\chi x_2}{R})]d^3x_2\nonumber \\
&&= {}\frac{g^2}{4\pi}N^2\int[j^2_0(\frac{\chi X}{R})-j^2_1(\frac{\chi X}{R})]d^3X\times\nonumber \\
&& {}\int[j^2_0(\frac{\chi |\bf X-\bf Y|}{R})-j^2_1(\frac{\chi
|\bf X-\bf Y|}{R})]\frac{e^{-M_\sigma Y}}{Y}D^3Y,
\end{eqnarray}
and
\begin{eqnarray}
&E_{qpq}&=\frac{g^2}{4\pi}4N^2\int j_0(\frac{\chi x_1}{R})j_1(\frac{\chi x_1}{R})({\mbox{\boldmath $\sigma_1$}}\cdot{\bf x}_1)d^3x_1\times\nonumber \\
&& {}\int\frac{e^{-M_p |{\bf x}_1-{\bf x}_2|}}{|{\bf x}_1-{\bf x}_2|}j_0(\frac{\chi x_2}{R})j_1(\frac{\chi x_2}{R})({\mbox{\boldmath $\sigma_2$}}\cdot{\bf x}_2)d^3x_2\nonumber \\
&&= {}\frac{g^2}{4\pi}4N^2\int j_0(\frac{\chi X}{R})j_1(\frac{\chi X}{R})({\mbox{\boldmath $\sigma_1$}}\cdot\hat{X})d^3X\times\nonumber \\
&& {}\int j_0(\frac{\chi |\bf X-\bf Y|}{R})j_1(\frac{\chi |\bf
X-\bf Y|}{R})({\mbox{\boldmath $\sigma_2$}}\cdot\frac{\bf X-\bf
Y}{|\bf X-\bf Y|})\frac{e^{-M_p Y}}{Y}d^3Y.
\end{eqnarray}
With the transformation, one can easily find that the formal
singularity is automatically cancelled in the spherical
integrations. It is interesting to discuss the integral range of
Y, which is related to the polar angle $\theta$ between ${\bf X}$
and ${\bf Y}$. Its minimum is zero, and its maximum is $f(\theta)$
with
\begin{eqnarray}
f(\theta)=\sqrt{X^2\cos^2\theta+R^2-X^2}-Xcos^2\theta.
\end{eqnarray}
Then we can write down the final formulation of interacting energy
due to one-meson-exchange, which corresponds to the
intermediate-diatnce interaction:
\begin{eqnarray}
E_{q\sigma q}&=&\frac{g^2}{4\pi}N^2_1\frac{N^2_2}{2}\int^R_0[j^2_0(\frac{\chi X}{R})-j^2_1(\frac{\chi X}{R})]X^2dX\times\nonumber \\
&& {}\int^\pi_0\sin\theta
d\theta\int^{f(\theta)}_0[j^2_0(\frac{\chi |\bf X-\bf
Y|}{R})-j^2_1(\frac{\chi |\bf X-\bf Y|}{R})]e^{-M_\sigma Y}YdY,
\end{eqnarray}
and
\begin{eqnarray}
E_{qpq}&=&\frac{g^2}{4\pi}4N^2_1\frac{N^2_2}{2}({\mbox{\boldmath
$\sigma_1$}}\cdot{\mbox{\boldmath $\sigma_2$}})\int^R_0
j_0(\frac{\chi X}{R})j_1(\frac{\chi
X}{R})X^2dX\times\nonumber \\
&& {}\int^\pi_0\sin\theta d\theta\int^{f(\theta)}_0 j_0(\frac{\chi
|\bf X-\bf Y|}{R})j_1(\frac{\chi |\bf X-\bf Y|}{R})\frac{X}{|\bf
X-\bf Y|}e^{-M_p Y}YdY.
\end{eqnarray}
Below, we will turn into
numerical computations.\\

\noindent{III. Numerical Results}

As discussed above, we hope to take both short-distance and
intermediate-distance interaction into account to get a more
complete physics picture of strong interaction among quarks in
hadrons. Namely, quark-gluon coupling and quark-meson coupling are
considered simultaneously. In this work, we still follow the
general strategy given by the MIT bag model, namely, summing over
the contributions of the short-distance and intermediate-distance
interactions which are discussed in last subsection, as well as
the bag vacuum energy $\frac{4}{3}\pi R^3B$ and zero point energy
$\frac{-Z_0}{R}$ to constitute the total energy and then
differentiate it with respect to the bag radius to obtain an
effective bag radius. Substituting the radius into the expression
of the energy, we obtain the formulation of the baryon mass
spectra. Then we need to fix the concerned parameters by fitting
data. In this new version, there are four parameters to be fixed,
$\alpha_c$, $\alpha_M$, $B$ and $Z_0$.

In the original paper on the bag model, the authors gave two sets
of $B$ and $Z_0$, here we only choose one set. We tried to vary
the values of $B$ and $Z_0$ near the original ones and to see if
we can fit the spectra. We find that it is hard to get
satisfactory results no matter how we change these two parameters.

The new parameter $\alpha_M$ which did not exist in the original
version of the MIT bag model, is to be fixed by fitting the rich
spectra of baryon octet and decuplet. In the potential model,
Glozman et.al. obtained a value for the quark-meson coupling
constant \cite{Glo}.  Since the short-distance and
intermediate-distance interactions have different algebraic
structures, the corresponding effective couplings are irrelevant,
so that in principle there may be several possible choices. But we
find that only certain combinations can best fit the data. It is
hinted that the value of $\alpha_M$ achieved by Glozman et.al
\cite{Glo} in the potential model may be appropriate for the
spectra evaluation, we would choose the value of $\alpha_M$ close
to that of \cite{Glo}. By varying their values, we obtain the best
fit to the spectra. Then we can compare our result with theirs. We
finally find that a combination of parameters:
$B^\frac{1}{4}=0.145$, $Z_0=1.84$, $\alpha_c=0.398$,
$\alpha_M=0.545$ can well accommodate the experimental spectra of
baryons.

We find our result about $\alpha_M$ which is a phenomenological
coupling constant between quark and chiral bosons is consistent
with that obtained by Glozman et al. in the potential model. With
the parameters, the fitted results are shown in the following
label:
\begin{center}
\begin{tabular}{|c||c|c|c|c|c|} \hline
$Baryon$ & $content$  & $M_{B}(exp)(GeV)$ & $M_{B}(our)(GeV)$ & $R(GeV^{-1})$ & $M_{B}(\cite{Jaffe})(GeV)$\\
\hline \hline
$P$&uud&$0.938$&0.938&5.0&0.938\\
\hline
$\Lambda$&uds&$1.116$&1.105&5.0&1.105\\
\hline
$\Sigma^+$&uus&$1.189$&1.180&5.1&1.144\\
\hline
$\Xi^0$&uss&$1.321$&1.319&5.0&1.289\\
\hline
$\Delta^+$&uud&$1.236$&1.253&5.6&1.233\\
\hline
$\Sigma^*$&uus&$1.385$&1.391&5.5&1.382\\
\hline
$\Xi^*$&uss&$1.533$&1.535&5.4&1.529\\
\hline
$\Omega^-$&sss&$1.672$&1.684&5.4&1.672\\
\hline
\end{tabular}

\end{center}

\centerline{Table 1. The concerned results, as a comparison we
list the results }

\centerline{given in the earlier work for the MIT bag model.}

\vspace{0.5cm}

\noindent{IV. Qualitative and semi-quantitative study on
three-body interactions}

Besides the one-gluon and one-meson exchange between two quarks,
there also could be interactions among all three quarks in nucleon
via a triple gluon vertex or $\sigma\pi\pi$ coupling vertex. Such
coupling processes are three-body interactions. For triple gluon
coupling case, as argued in \cite{wang}, the interacting energy is
proportional to:
\begin{equation}
f^{abc}\epsilon_{ijk}\epsilon_{i'j'k'}\lambda^a_{ii'}\lambda^b_{jj'}\lambda^c_{kk'},
\end{equation}
analysis indicates that due to the requirement of color singlet
for hadrons, this contribution of such an interaction is null.

Then we turn to the $\sigma\pi\pi$ case.

Following the procedure before, there is a product of three
propagators:
\begin{eqnarray}
\frac{1}{{\bf l^2}+m^2_\sigma}\times\frac{1}{{\bf
q^2}+m^2_\pi}\times\frac{1}{{\bf p^2}+m^2_\pi},
\end{eqnarray}
where {\bf l}, {\bf q}, {\bf p} are the 3-momentum of $\sigma,
\pi, \pi$ respectively. By the momentum conservation, it is:
\begin{eqnarray}
\frac{1}{{\bf (p+q)^2}+m^2_\sigma}\times\frac{1}{{\bf
q^2}+m^2_\pi}\times\frac{1}{{\bf p^2}+m^2_\pi}.
\end{eqnarray}

To get the final form in the coordinate space, we make a
Fourier-transformation first on {\bf q}:
\begin{eqnarray}
\int Ae^{i{\bf q\cdot r'}}\frac{d^3q}{(2\pi)^3}&=&\frac{1}{4\pi r'}[\frac{e^{-m_\pi r'}}{(p-i(m_\sigma-m_\pi))(p+i(m_\sigma+m_\pi))}\nonumber \\
&& {}+\frac{(-p+im_\sigma)e^{-ipr'-m_\sigma
r'}}{im_\sigma(p-i(m_\sigma-m_\pi))(p-i(m_\sigma+m_\pi))}]
\end{eqnarray}
where
\begin{eqnarray}
A=\frac{1}{{\bf (p+q)^2}+m^2_\sigma}\times\frac{1}{{\bf
q^2}+m^2_\pi}
\end{eqnarray}
Then on {\bf p}:
\begin{eqnarray}
V_{prop} &=& \int \frac{B}{{\bf p^2}+m^2_\pi}e^{i{\bf p\cdot
r}}\frac{d^3p}{(2\pi)^3} \nonumber \\
&=&\frac{1}{4\pi r}\frac{1}{4\pi r'}[\frac{e^{-m_\pi(r+r')}}
{(m_\sigma+2m_\pi)(m_\sigma-2m_\pi)}
-\frac{(m_\sigma-m_\pi)e^{-(m_\sigma-m_\pi)r'}e^{-m_\pi r}}
{(m_\sigma-2m_\pi)m^2_\sigma}\nonumber \\
&& -2\frac{(m_\sigma-m_\pi)e^{-m_\pi r'}e^{-(m_\sigma-m_\pi)r}}
{(m_\sigma-2m_\pi)m^2_\sigma} +\frac{(m_\sigma+m_\pi)e^{m_\pi
r'}e^{-(m_\sigma+m_\pi)r}}{(m_\sigma+2m_\pi)m^2_\sigma}],
\end{eqnarray}
where
\begin{eqnarray}
B&=&[\frac{e^{-m_\pi r'}}{(p-i(m_\sigma-m_\pi))(p+i(m_\sigma+m_\pi))}\nonumber \\
&& {}+\frac{(-p+im_\sigma)e^{-ipr'-m_\sigma
r'}}{im_\sigma(p-i(m_\sigma-m_\pi))(p-i(m_\sigma+m_\pi))}].
\end{eqnarray}
Eq.(26) provides us the "formal" propagator in $\sigma\pi\pi$
coupling process. Then the interacting energy caused by the
three-meson coupling is written as:
\begin{eqnarray}
E_{\sigma\pi\pi}=-g_{\sigma\pi\pi}\int
\overline{\psi}'_1g\psi_1V_{prop}\overline{\psi}'_2g\gamma^5\psi_2\overline{\psi}'_3g\gamma^5\psi_3d^3xd^3yd^3z
\end{eqnarray}
where $g_{\sigma\pi\pi}$ is the coupling vertex calculated in
\cite{Schu,Zhuang}:
\begin{eqnarray}
g_{\sigma\pi\pi}\sim2GeV
\end{eqnarray}
The complete calculation is very difficult, so we would estimate
its order of magnitude in a simplified scenario where only the
"breathing mode" is considered. It means that although the three
quarks can reach any point in the bag, the relative angles among
them remains at $120^{\circ}$, and the relative spacial distances
among them are the same all the time. Then we can calculate the
three-body interacting energy with the simplified scenario.
Numerical integration on computer indicates that the upper limit
of $E_{\sigma\pi\pi}\sim0.005GeV$, which is much smaller than that
by the one-gluon and one-meson exchanges. Therefor with the
present experimental accuracy, we need not take this contribution
into account at all. Although this picture is rough, one can
expect that it can at
least give the right order of magnitude of such interaction.\\

\noindent{V. Conclusion and Discussions}

Quarks and gluons are confined inside hadrons by strong
interaction which is described by QCD. Due to the asymptotic
freedom, the quarks which are close to each other are
approximately free of interaction and it is the basic point of the
MIT bag model, in which the quarks obey the Dirac equation for
free fermions with a bag boundary condition. Indeed it is the
non-perturbative QCD effects which correspond to the long-distance
interaction and bind quarks into a hadron. Even though the
potential model looks quite different from the MIT bag model,
basically, they are somehow equivalent and just the linear or some
other confinement potentials replace the bag boundary in the MIT
bag model. Since so far, there lacks a reliable way to approach
the non-perturbative QCD and  a unique picture from quark-gluon
degrees of freedom to the hadron phase cannot be derived from any
underlying theory.

It is generally believed that the one-gluon exchange which
obviously represents the leading order in QCD, corresponds to the
short-distance interaction between quarks. In the MIT bag model,
it is accounted as a correction to the total energy and its
contribution is evaluated in perturbation method. On other side,
the chiral boson-exchanges are also supposed to contribute an
intermediate-distance interaction which also plays a role to bind
quarks in hadrons. It is argued \cite{Win,Geor1,Geor2,yanmulin}
that the chiral bosons are also interaction agents between quarks
and correspond to degrees of freedom which are independent of
gluons at the energy region of $\Lambda_{QCD}$. Moreover, some
authors\cite{Gui,Fle,Sai,Blu} claim that the gluon-exchange can be
dropped out due to the asymptotic freedom and only the
chiral-boson-exchanges apply or at least dominate.  It seems to
contradict to the approach of the MIT bag model and this problem
concerns the fundamental physics picture, so is worth careful
investigation, i.e. if the two pictures are consistent. In this
work, we just include the contribution from both gluon-exchange
and chiral-boson-exchange to the total energy and see if the
results make sense. The purpose of this work is not to gain any
better phenomenological predictions which can be experimentally
tested, but tries to clarify the physics picture and see if one
can accommodate the short-, intermediate- and long-distance
interactions in a unique framework. Definitely, the MIT bag model
among various models for hadron spectra provides an ideal place to
study this subject.

In this work, we only consider the exchanges of $\sigma$ and
$\pi{(\pm,0)}$ and ignore the contributions from exchanges of
vector mesons because they are much heavier than the scalar and
pseudoscalar mesons. Since we obtain the corresponding parameters
by fitting data, there exist certain errors coming from
experimental measurement, especially for the heavier members of
the baryon octet and decuplet. Moreover, we need a set of
parameters and the spectra of well measured baryons can determine
all of them. In particular, we determine the effective quark-meson
coupling and compare the value with that obtained in potential
model. We find that our result is consistent with the given by
Glozman et al \cite{Glo}. For being more confident with the
results, we also roughly estimate the three-body interactions
among the three valence quarks. The Lie algebra indicates that the
interaction via the three-gluon vertex is null due to the color
singlet requirement for baryons, whereas the interaction via
$\sigma\pi\pi$ vertices can result in non-zero contributions.
Since a complete calculation is extremely difficult, we only use a
simplified picture, namely only the breathing mode is considered,
to estimate the order of magnitude of such three-body interaction.
We find that this contribution is much smaller than the two-body
interactions and generally can be safely ignored from practical
calculations. We admit that because this treatment is very
simplified, the result may deviate from the real value, however,
we believe that the order of magnitude must be correct and the
qualitative conclusion about the size of the three-body
interaction is close to reality.

By our numerical results, we can conclude our findings as
following.

Letting the differentiation of the total energy which is a
function of the bag radius $R$ with respect to $R$ be zero, we
obtain the radius $R$, and the total energy with this $R-$value is
a minimum and  supposed to be the real mass of the baryon. The
expression of the total energy includes contributions from the
energy-eigenvalues of free quarks which corresponds to the zeroth
order of strong Hamiltonian, single-gluon-exchange,
chiral-meson-exchange, the vacuum pressure term ${4\over 3}\pi
R^3B$ and the zero-point energy ${-Z_0\over R}$. By re-fitting the
well-measured baryon spectra, we have obtained the concerned
parameters which are listed in last section. Comparing with the
parameter values obtained in the early works about the MIT bag
model where only one-gluon-exchange was considered, the best fit
to the data shows that the contribution from the
intermediate-distance effect induced by chiral-meson-exchanges can
be as large as 40\% of that from short-distance effect induced by
gluon-exchange, while the $B-$ and $Z_0-$values remain unchanged.
This fact indicates that inclusion of the chiral-boson exchanges
which induce the intermediate-distance interaction, changes the
effective coupling $\alpha_c$ of the previous work where only
short-distance interaction was considered, but does not affect the
long-distance interaction which is manifested by $B$ and $Z_0$. As
a conclusion, in a complete picture, both short-distance and
intermediate-distance interactions should be involved, however, if
one uses an effective coupling for either short-distance
(gluon-exchange) or intermediate-distance (chiral-boson-exchange),
the phenomenology is the same, but the effective coupling would
have different values.

Indeed, in this work, we ignore contributions from vector mesons
because of their heavier masses and couplings and also omit the
three-body interaction because of its smallness in comparison with
the two-body interactions. This treatment may bring up certain
errors definitely, but should not influence our qualitative
conclusion.

Even though there is no any difference for evaluating spectra of
baryons as long as one uses right effective coupling in either of
the two scenarios, namely only considers one type of exchanges,
gluon or chiral bosons, the difference may manifest itself when
evaluating some dynamical processes, such as decays. We will
further investigate these processes in our later works.\\

\noindent{Acknowledgements:}

This work is partly supported by the National Natural Science
Foundation of China.

\end{document}